\documentclass[prl,twocolumn,showpacs,preprintnumbers,amsmath,amssymb,floatfix]{revtex4}
\usepackage{graphicx}
\usepackage{dcolumn}
\usepackage{bm}
\usepackage{epsfig}

\begin{document}

\newcommand{\bec}{\begin{center}}
\newcommand{\ec}{\end{center}}
\newcommand{\be}{\begin{equation}}
\newcommand{\ee}{\end{equation}}
\newcommand{\beqn}{\begin{eqnarray}}
\newcommand{\eeqn}{\end{eqnarray}}
\newcommand{\bet}{\begin{table}}
\newcommand{\ent}{\end{table}}
\newcommand{\bib}{\bibitem}



\title{
 L\'evy-flight intermixing: 
anomalous nanoscale diffusion in Pt/Ti
}

\author{P. S\"ule, M. Menyh\'ard} 
\affiliation{
  Research Institute for Technical Physics and Material Science,
www.mfa.kfki.hu/$\sim$sule, sule@mfa.kfki.hu\\
Konkoly Thege u. 29-33, Budapest, Hungary\\
}

\date{\today}

\begin{abstract}
Probing the anomalous nanoscale intermixing using molecular dynamics (MD) simulations
in Pt/Ti bilayer we reveal the superdiffusive nature of
interfacial atomic transport. 
It is shown that the Pt atoms undergo anomalous atomic transport across the
anisotropic interface of Pt/Ti with suprisingly high rates which
can be characterized as L\'evy flights.
L\'evy flight is not a unique phenomenon in nature, however, no such events
have been reported yet for bulk interdiffusion.
In particular,
the low-energy ($0.5$ keV) ion-sputtering induced transient enhanced intermixing has been studied by MD simulations.
{\em Ab initio} density functional calculations have been used to check and reparametrize the employed heteronuclear interatomic potential.
The L\'evy-intermixing behavior explains the high diffusity tail in the concentration profile obtained
by 
Auger electron spectroscopy depth profiling (AES-DP) analysis in Pt/Ti bilayer (reported in ref.: P. S\"ule, {\em et al.},
J. Appl. Phys., {\bf 101}, 043502 (2007)).

\pacs{ 05.40.Fb, 68.35.Fx, 02.50.Ey, 66.30.Jt, 61.80.Jh, 66.30.-h, 
}
\end{abstract}
\maketitle

 Anomalous diffusion has long been known since the classical Richardson's study on turbulent diffusion \cite{Richardson}.
 Since than anomalously large diffusion rates have been observed in various
systems under different conditions \cite{anomalous,Levy,LF,Bouchaud,Shraiman,NanoLett}.
The common feature of the observed enhanced diffusion rates
are the
nonlinear growth of the mean squared displacement in the course of time 
leading to a power-law time scaling pattern \cite{anomalous}.
Hence we are faced with non-Brownian motion of particles with non-Gaussian
diffusion rates giving rise to anomalous behaviors which are termed as
{\em strange kinetics} \cite{Shlesinger}.
Superdiffusion takes place with L\'evy flight characteristics (LF, L\'evy jump diffusion with broad spatial jump distributions) above the Brownian motion regime \cite{anomalous}.
In paticular, LF behaviour has been found in
turbulent mixing in fluids (review in refs. \cite{anomalous,Bouchaud,Shraiman})
or in elongated micelles \cite{Bouchaud}, for photon trajectories in incoherent atomic radiation trapping \cite{Pereira} and it became clear that 
L\'evy processes can be important in the analysis of chaotic dynamics \cite{Levy}.
The research of
LF behaviors covers many disciplines of natural and social sciences hence
this topic is of high interest \cite{anomalous,Levy,Levy2,LF,Shraiman,Klafter}.
Our primary purpose is to
present experimental and computational evidences that
the nanoscale transient enhanced intermixing and diffusion, a common phenomenon in many nanostructured materials,
could also be fitted in the category of L\'evy flight superdiffusion.

 There are a growing number of evidences emerged 
in the last decades
that anomalous nanoscale broadening of interfaces or high diffusity tail in the impurity concentration profile
occur during ion-irradiation or sputtering \cite{Abrasonis,Weber,Cardenas,Nieveen,Sule_JAP07}, sputter deposition and thin film growth \cite{Buchanan,Sule_JCP}.
The anomalous nanoscale bulk diffusional effects
could be due to still not clearly established atomistic accelerative effects leading to anomalously fast and possibly athermal atomic transport (long range diffusion) at interfaces or during impurity diffusion
\cite{Abrasonis,Weber,Cardenas,Nieveen,Sule_JAP07,Buchanan,Avasthi}.

 In the present work, using computer atomistic simulations,
   the recently observed enhanced intermixing in a nanoscale bilayer film/substrate system (Pt/Ti) reported in ref. \cite{Sule_JAP07} is attempted to interpret as a L\'evy-superdiffusive atomic transport
process in the bulk. 
Superdiffusion in the nanoscale has only been reported until now on solid surfaces
\cite{Luedtke,Amar} and no reports have been found
for bulk superdiffusion in the solid state of matter.


 Classical molecular dynamics simulations have been used to simulate the ion-sputtering
(repeated low-energy ion bombardments of the film)
induced atomic intermixing at the Pt/Ti interface
using the PARCAS code \cite{Nordlund_ref}.
Here we only shortly summarize the most important aspects.
A variable timestep
and the Berendsen temperature control is used to maintain the thermal equilibrium of the entire
system at $300$ K (see refs. at \cite{Sule_JAP07}). 
The detailed description of other technical aspects of the MD simulations are given in refs. \cite{Nordlund_ref} and details specific to the current system in recent
communications \cite{Sule_JAP07,Sule_PRB05,Sule_NIMB04}.

 Our primary purpose is to simulate the conditions occur during ion-sputtering \cite{Sule_JAP07}
and Auger electron spectroscopy depth profiling analysis (AES-DP) \cite{Sule_JAP07}
using molecular dynamics simulations.
Following our previous work \cite{Sule_JAP07}
we ion bombard the film of the bilayers Pt/Ti and Ti/Pt (film/substrate systems)
with 0.5 keV Ar$^+$ ions repeatedly (consequtively) with a time interval of 10-20 ps between each of
the ion-impacts at 300 K
which we find
sufficiently long time for most of the structural relaxations and the termination of atomic mixing, such
as sputtering induced intermixing (IM) \cite{Sule_NIMB04}.
Since we focus on the occurrence of transient intermixing atomic transport processes,
the relaxation time of $10-20$ ps should be appropriate for getting adequate information
on transient enhanced intermixing.
Pair potentials have been used
 for the interaction of the Ar$^+$ ions with the metal atoms derived using
{\em ab initio} density functional calculations.
 The initial velocity direction of the
impacting ion was $10$ degrees with respect to the surface of the film (grazing angle of incidence)
to avoid channeling directions and to simulate the conditions applied during ion-sputtering \cite{Sule_JAP07}. 
The impact positions have been randomly varied on the surface of the film/substrate system and the azimuth angle $\phi$ (the direction of the ion-beam).
In order to simulate ion-sputtering a large number of ion irradiation are
applied using script governed simulations conducted subsequently together with analyzing
the history files (movie files) in each irradiation steps.
The impact positions of the $400$ ions are randomly distributed
over a $20 \times 20$ \hbox{\AA}$^2$ area on the surface.

\begin{table}[t]
\caption
{
The semiempirical parameters used in the tight binding interatomic potential for Pt/Ti \cite{Sule_JAP07,CR}
}
\begin{ruledtabular}
\begin{tabular}{cccccc}
& $\xi$ & q & A & p & $r_0$  \\
\hline
 Ti  & 1.416  & 1.643  & 0.074 & 11.418 & 2.95 \\
 Pt  & 2.695 & 4.004  & 0.298 & 10.612 & 2.78   \\
 Ti-Pt & 4.2 & 2.822 & 0.149  & 11.015 & 2.87   \\
\hline
\end{tabular}
\end{ruledtabular}
\footnotetext[1]{
The parameters for Ti and Pt have been given by Cleri and
Rosato \cite{CR}.
The parameters of the crosspotential have been obtained
by fitting the interpolated crosspotential given in
ref. \cite{Sule_JAP07} to {\em ab initio} diatomic calculations.
The preexponential parameter $\xi$ has been fitted to the {\em ab initio} curve
(see also in ref. \cite{Sule_JCP2}).
The notations used for the parameters are the same as given in refs.
\cite{Sule_JAP07,Sule_JCP,CR}.
}
\label{table}
\end{table}

 We used a tight-binding many body potential
on the basis of the second moment approximation (TB-SMA) to the density of states \cite{CR}, to describe interatomic interactions.
In ref. \cite{Sule_JAP07}
it has been shown that the TB-SMA potential gives the reasonable description of IM in Pt/Ti
and gives interfacial broadening comparable with AES-DP measurements.
 The crosspotential energy has been calculated for the Ti-Pt dimer
  using {\em ab initio} local spin density functional calculations \cite{G03} together with quadratic convergence self-consistent field (SCF) method.
The G03 code is well suited for molecular calculations, hence
it can be used for checking pair-potentials.
We used the Perdew-Burke-Ernzerhof (PBE) gradient corrected exchange-correlation potential \cite{PBE}.
We find that the interpolated TB-SMA potential \cite{Sule_JAP07}
nearly perfectly matches the ab initio one hence we are convinced
that the TB-SMA model accurately describes the heteronuclear
interaction in the Ti-Pt dimer.
The employed parameter set is given in Table 1.

 The crossectional computer animations of simulated ion-sputtering can be seen in our web page \cite{web}.
  The cartoons of the simulation cells (crossectional slabs as a 3D view)
can be seen in ref. \cite{Sule_JAP07} which show the strong intermixing at the interface
in Pt/Ti and a much weaker mixing in Ti/Pt.
\begin{figure}[hbtp]
\begin{center}
\includegraphics*[height=5.5cm,width=7cm,angle=0.]{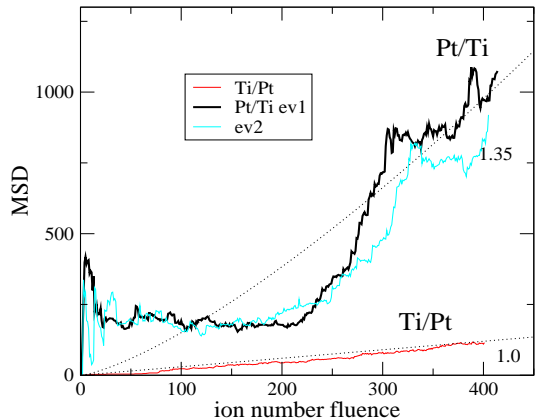}
\caption[]{
The simulated mean square of intermixing atomic displacements $\langle R^2 \rangle_z$ (MSD, variance in $\hbox{\AA}^2$/particle) in Pt/Ti and  Ti/Pt as a function of the
ion-fluence (the number of ion bombardments, $N_{ion}$) obtained during the ion-sputtering of these bilayers at 500 eV ion energy (results are shown up to $\sim 400$ ions).
Results from two independent simulations are shown for Pt/Ti.
Fits to the variance of various curves are also given with dotted lines which
is obtained as $\langle R^2 \rangle_z \approx N^{\alpha}$.
The obtained values of $\alpha$ are also shown.
}
\label{R2}
\end{center}
\end{figure}
  In Fig ~\ref{R2} the evolution of the mean-square of atomic displacements (MSD)  (the variance of displacements)
$\langle R^2 (t) \rangle_z= \sum_j^{N_{ion}} \frac{1}{N_{atom}^j} \sum_i^{N_{atom}^j} [{\bf r_i^j}(t)-{\bf r_i}(t=0)]^2$,
of all intermixing atoms                                                           obtained by molecular dynamics simulations, where (${\bf r_i}^j(t)$ is the position
 vector of atom 'i' at time $t$, $N_{atom}^j(t)$ is the total number of intermixing atoms in the $j$th irradiation step
included in the sum), can be followed as a function of the ion fluence (the number of ions $N_{ion}$).
Lateral components ($x,y$) are excluded from
$\langle R^2 \rangle_z$ and only contributions from IM atomic displacements perpendicular to the layers
are included ($z$ or depth components). We follow during simulations the time evolution
of $\langle R^2 \rangle_z$ which reflects the atomic migrations through the interface (no other
atomic transport processes are included).

  In Fig. ~\ref{R2} we present
$\langle R^2 \rangle_z$ as a function of the number of ion impacts $N_{ion}$ (ion-number fluence).
$\langle R^2 \rangle_z (N_{ion})$ corresponds to the final value of
$\langle R^2 \rangle_z$
obtained during the $N_{ion}$th simulation. The final relaxed structure of the simulation of the
$(N_{ion}-1)$th ion-bombardment is used as the input structure for the $N_{ion}$th ion-irradiation.
The
asymmetry of
 mixing can clearly be seen when $\langle R^2 \rangle_z (N_{ion})$
and the depth profiles given in ref. \cite{Sule_JAP07} are compared in 
Ti/Pt and in Pt/Ti.
The computer animations of the simulations \cite{web} together with the plotted broadening values
at the interface in 
ref. \cite{Sule_JAP07} also reveal the stronger
IM in Pt/Ti.
 Moreover we find the strong divergence of $\langle R^2 \rangle_z$ from linear scaling
for Pt/Ti and a much weaker nonlinear scaling has been found for Ti/Pt.
 As it has already been shown in ref. \cite{Sule_JAP07}
AES-DP found
\begin{figure}[hbtp]
\begin{center}
\includegraphics*[height=4cm,width=4.2cm,angle=0.]{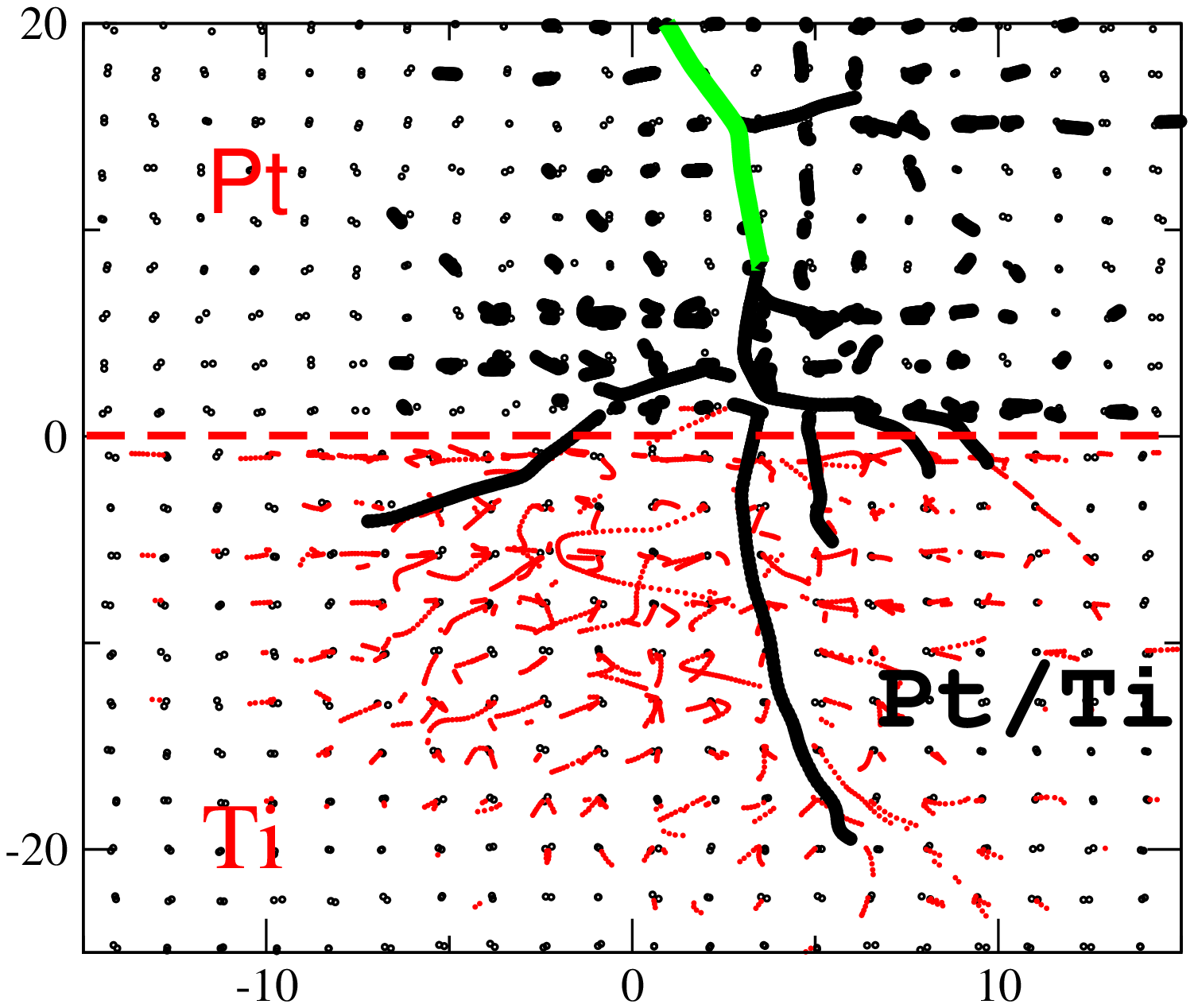}
\includegraphics*[height=4cm,width=4.2cm,angle=0.]{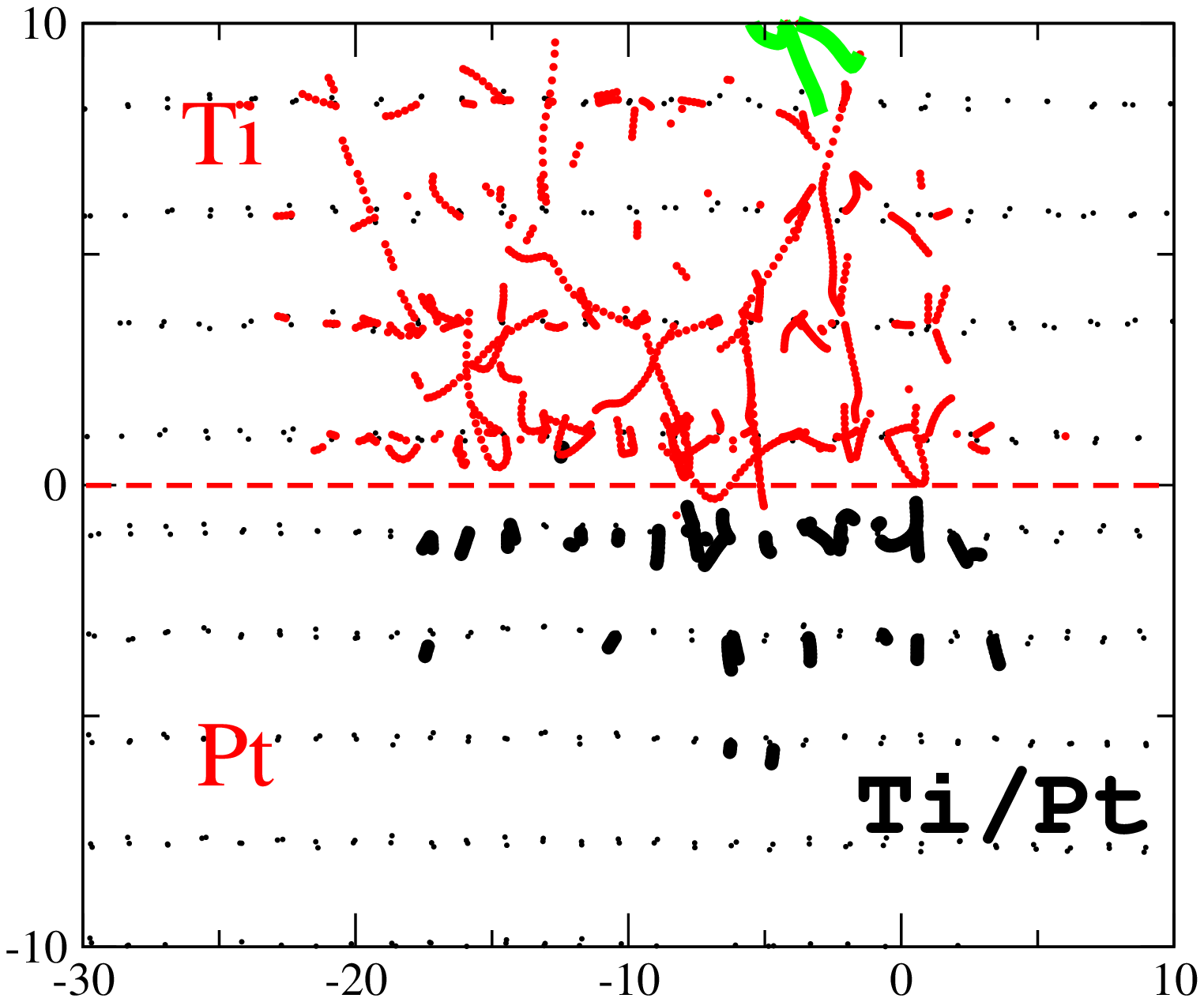}
\caption[]{
The normal to the surface crossectional view of collisional displacement cascades
with atomic trajectories
(crossectional slab cut in the middle of the simulation cell)
in Pt/Ti and in Ti/Pt.
The right and left panels correspond to Pt/Ti and to Ti/Pt, respectively.
The positions of the energetic particles of which the kinetic energy
is larger than $\sim 1$ eV
are collected during
typical single ion impact events
at $500$ eV ion energy.
The ions have been initialized $10$ $\hbox{\AA}$ above the interface.
The vertical axis corresponds to the depth position given in $\hbox{\AA}$.
The position at $z=0$ is the depth position of the interface with a horizontal dashed line 
and the initial positions of the atoms (before irradiation) are also shown.
}
\label{flight}
\end{center}
\end{figure}
 a relatively weak IM in Ti/Pt (the interface broadening $\sigma \approx 20$ $\hbox{\AA}$) while
an unusually high
IM occurs in the Pt/Ti bilayer ($\sigma \approx 70$ $\hbox{\AA}$).
 In order to clarify the mechanism of intermixing and to understand how much
the nanoscale interfacial mass-anisotropy influences IM,
  simulations have been carried out with atomic mass ratio $\delta=m_{Pt}/m_{Ti}$ (where $m_{Pt}$ and $m_{Ti}$ are the atomic masses) is artificially set to 
$\delta \approx 1$ (mass-isotropy).
We find that the magnitude of IM is strongly sensitive to $\delta$ at
mass-anisotropic, called $\delta$-interfaces.
The corresponding animation can be also be seen at \cite{web}.
We reach the conclusion that the mass-effect is robust and the magnitude of IM is
weakened significantly in artificially isotropic Pt/Ti in agreement with
our earlier finding \cite{Sule_PRB05,Sule_NIMB04}.
The huge difference in IM
between Pt/Ti and
Ti/Pt can be understood as
the effect of $\delta$-inversion on IM.
Actually the system undergoes the transition in the asymptotics of $\langle R^2 \rangle
\propto t^{1.35} \rightarrow \langle R^2 \rangle \propto t^{1.0}$
(from nonlinear dynamics to linear) when the
mass anisotropy is inverted (the film and the substrate is interchanged).

 To further test mass-effect on IM, we carried out simulations
for the Pt/Ti system in which the atomic masses have been interchanged (Ti possesses the atomic mass of Pt and vice versa) setting in
an artificial mass ratio (the inverse of the normal one)
while keeping all the other parameters unchanged.
Note, that we keep the interatomic potentials, only the atomic mass ratio is inverted.
We find that this artificial setup of atomic masses results in the suppression
of IM in Pt/Ti with inverted $\delta$.
These findings together with our AES measurements (with the long-range tail shown in ref. \cite{Sule_JAP07})
confirms our recent results reported for
various bilayers in which a strong correlation has been obtained between
\begin{figure}[hbtp]
\begin{center}
\includegraphics*[height=4cm,width=4.2cm,angle=0.]{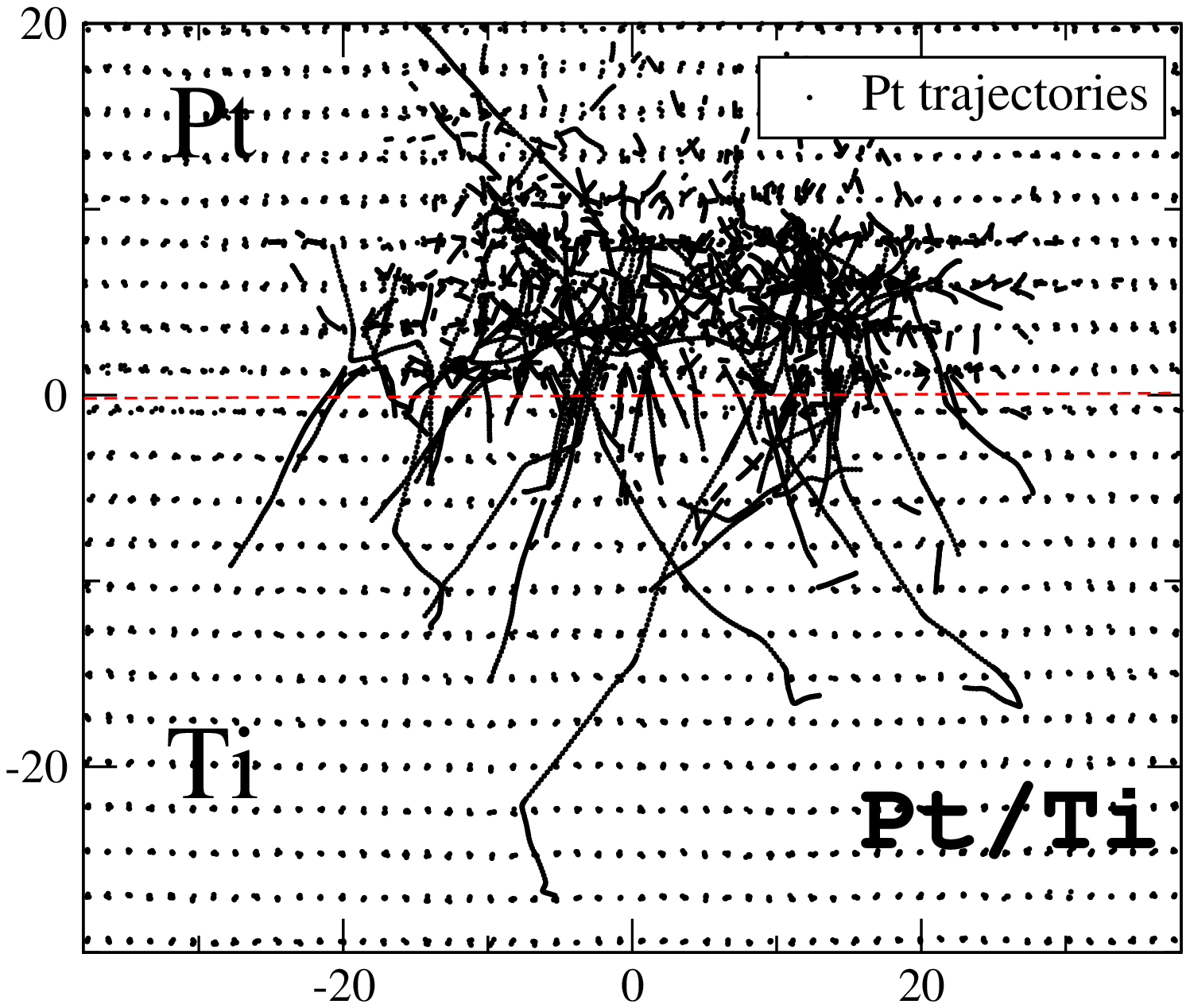}
\includegraphics*[height=4cm,width=4.2cm,angle=0.]{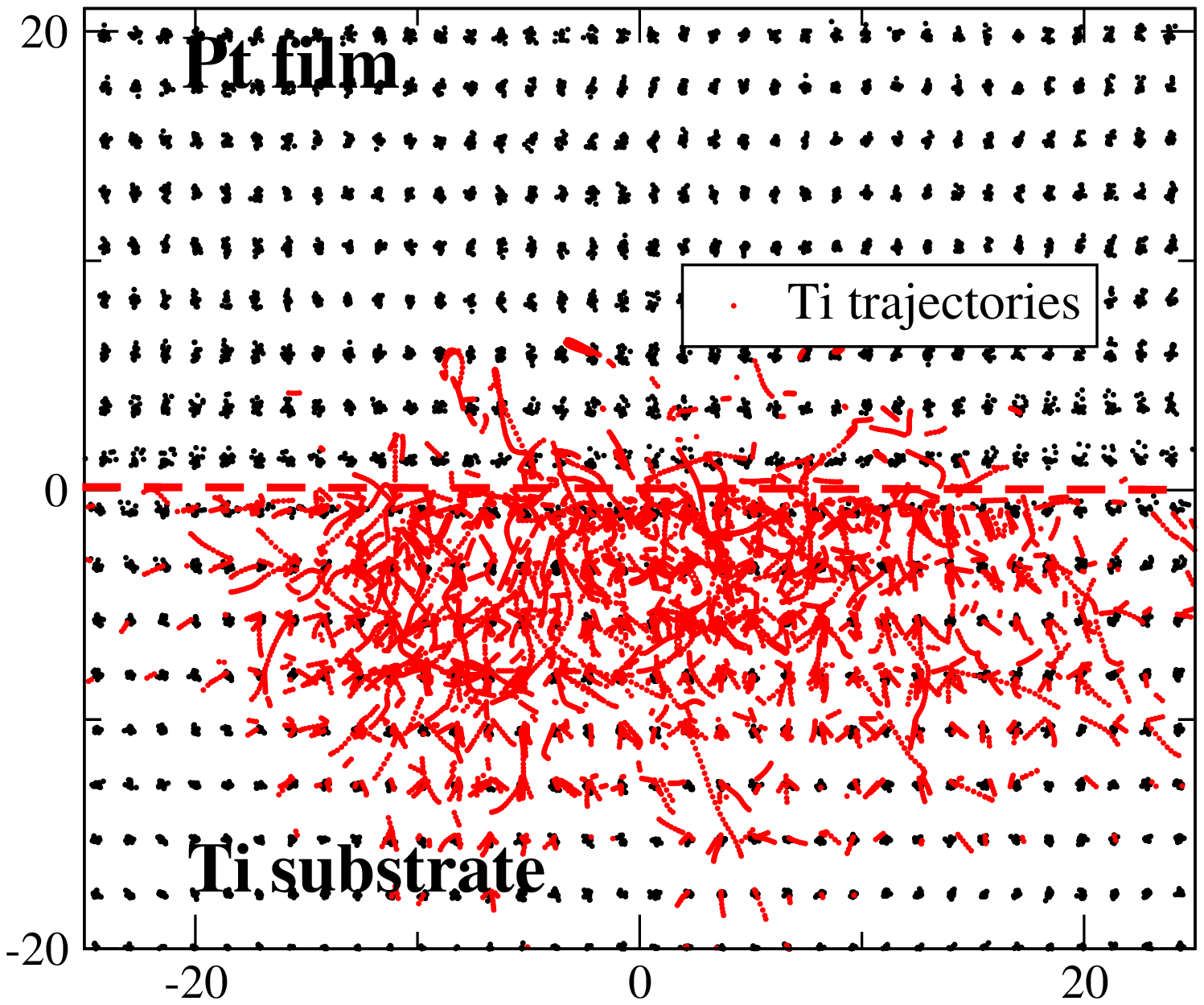}
\caption[]{
The normal to the surface crossectional view of 
trajectories of L\'evy flights
obtained during $50$ consequtive ion-impact events in Pt/Ti using 
$0.5$ keV ion energy for Pt (left panel). 
The atomic trajectories of hyperthermal Ti atoms are also shown
collected during the same irradiation events 
(panel on the right).
The atomic layers above the interface with depth position
$z > 0$ are Pt layers.
The positions of energetic atoms have been plotted which
have larger than $\sim 2$ eV kinetic energy.
The ions have been initialized $10$ $\hbox{\AA}$ above the interface.
The interface is at $z=0$ depth position (the increments are in $\hbox{\AA}$).
The initial positions of the atoms (before irradiation) are also shown.
}
\label{flights}
\end{center}
\end{figure}
the experimental and simulated mixing efficiencies and mass anisotropy in various metallic
bilayers
\cite{Sule_PRB05}.

  In Fig. ~\ref{R2} for Pt/Ti it can clearly be seen that
$\langle R^2 \rangle_z$ (MSD) grows nonlinearily with $N_{ion}$.
The horizontal axis is proportional to the time of ion-sputtering, hence
$\langle R^2 \rangle_z
\propto t^{\alpha}$ with $\alpha >1$.
\cite{anomalous}.
During thermally activated interdiffusion
$\langle R^2 \rangle$ scales
linearily with time
hence our system violates this temporal
behavior.
The $\langle R^2 \rangle_z \propto t^{\alpha}$, scaling,
where $\alpha \ge 1$,
used
to be
considered as the signature of
anomalous diffusion (superdiffusion)
in the literature \cite{anomalous,Levy,Levy2}.
In our case we find that $\alpha \approx 1.35 \pm 0.1$ exponent
fits the simulated curve
for Pt/Ti.
We find for Ti/Pt a linear behavior ($\alpha \approx 1.0 \pm 0.05$) hence no
superdiffusive features can be identified in this system.

  We would like to show that it might also be the case that
transient IM takes place in Pt/Ti which resembles in many respect
the L\'evy superdiffusive atomic transport processes known on solid surfaces \cite{anomalous}
.
Superdiffusion occur when the particle trajectories exhibit long
displacements (flights) termed L\'evy flights \cite{anomalous},
characterized by power-like asymptotical distribution
of the displacements of trajectories (and with heavy-tailed distribution) \cite{Levy}.
For these ballistic processes $\langle R^2 \rangle (t)$ is divergent with time.
Normal diffusion is characterized by the linear time evolution of $\langle R^2 \rangle$
in the long-time limit
and with Gaussian distribution \cite{Levy}.

  The obtained
results are in agreement
with the findings presented in ref. \cite{Sule_JAP07}.
However, in that paper
it has
not been realized
that
the fingerprint of superdiffusive
feature
of IM is detected by AES as a long-range diffusity tail in the concentration profile
at the Pt/Ti interface in the Pt/Ti bilayer.
No such tail occurs in the concentration profile of Ti/Pt as shown in ref. \cite{Sule_JAP07}.
Hence we conclude that the experimental fingerprint of L\'evy jumps could also be
detected as a high diffusity tail in the AES concentration profile of many
other anisotropic materials.


 The trajectories corresponding to L\'evy flights can be seen in the left panel in Fig. ~\ref{flight}.
In the panel of Pt/Ti in Fig ~\ref{flight} we can see the ballistic trajectories of intermixing hyperthermal Pt atoms which exhibit L\'evy flights with long trajectories through the
interface.
We find $8$ events out of $10$ which exhibit L\'evy flights.
No such trajectories can be seen in the panel of Ti/Pt in Fig. ~\ref{flight}
and in the other dozens of events generated (not shown).
The ions have been initialized during these single ion impact events
$10$ $\hbox{\AA}$ above the interface in order to place the range of
the ions directly in the depth position of the interface.
This way of direct deposition of the ion energy at the interface does not influence the
main physics what we find with simulations when ions have been initialized
at the surface.
This is because the first few tens of ions simply sputter remove the top layers
of the film and intermixing develops when the range of ions approaches the interface.
Fig. ~\ref{flights} depict us what we see in Fig. ~\ref{R2},
that L\'evy flights cause the nonlinear time scaling of $\langle R^2 \rangle_z$.
The initial kinetic energy of few of these particles can reach
the few tens of eV and which hyperthermal Pt atoms exhibit long trajectories.

  In the left panel of Fig. ~\ref{flights} we show all the flight trajectories of Pt atoms obtained
during a simulation with $50$ repeated ion impacts with $10$ ps time delay
between each of the events in Pt/Ti.
This figure clearly depicts us that L\'evy jumps of energetic Pt atoms
take place which boost intermixing.
Such kind of an interfacial broadening can also be called L\'evy interdiffusion.
In the right panel of Fig. ~\ref{flights} the trajectories of energetic Ti atoms are also shown 
obtained during the same simulation.
No L\'evy jumps can be seen for Ti atoms. These atoms not even move across the
interface except during few events.
In Ti/Pt no LF behavior can be found both for Pt and Ti atoms. 
Concerning the atomic mobility of the hyperthermal Pt particles,
some of these particles
can have a huge initial kinetic energy exceeding $100$ eV in few cases.
This could be due to the accelerative effect of head on collision of the ion or recoils with
few of the Pt atoms at the $\delta$-interface.
The Pt atoms intermix preferentially both in Ti/Pt and in Pt/Ti \cite{Sule_NIMB04}.
These rapidly migrating particles slow down to few eV within $0.5$ ps
and which energy regime persists up to few ps.
Hence a long range tail exists for the time distribution of
the atomic velocity during L\'evy jumps in Pt/Ti which is also a characteristic
feature of non-Brownian dynamics \cite{anomalous,Levy}.

 In conclusion, we reveal that
atomic intermixing might take place via L\'evy flights in the Pt/Ti film/substrate
bilayer upon low-energy ion bombardments.
This mechanism could be valid for many other interdiffusion processes in
which mass anisotropic interface is present.

This work is supported by the OTKA grants F037710
and K-68312
from the Hungarian Academy of Sciences.
The work has been performed partly under the project
HPC-EUROPA (RII3-CT-2003-506079) using
the supercomputing 
facility at CINECA in Bologna and at NIIF (Budapest).

\end{document}